\documentclass[a4paper,12pt,english,german]{article}

\begin{document}

\centerline{\bf The Ghostly Quantum Worlds}

\bigskip

\centerline{Miroljub Dugi\' c}

\centerline{Department of Physics, Faculty of Science, Radoja
Domanovi\' ca 12, 34000 Kragujevac, Serbia}

\centerline{Dejan Rakovi\' c}

\centerline{Faculty of Electrical Engineering, Bulevar kralja
Aleksandra 73, 11120 Belgrade, Serbia}

\centerline{Jasmina Jekni\' c-Dugi\' c}

\centerline{Department of Physics, Faculty of Science, Vi\v
segradska 33, 18000 Ni\v s, Serbia}

\centerline{Momir Arsenijevi\' c}

\centerline{Department of Physics, Faculty of Science, Radoja
Domanovi\' ca 12, 34000 Kragujevac, Serbia}

\bigskip

\centerline{\bf Abstract}

We present the foundations of a new emerging interpretation of
quantum theory bearing wide-range implications. Physical basis of
the interpretation is non-questionable yet relatively new--it
relies on the different structures (decompositions into parts,
subsystems) of the quantum Universe. We compare the mutually
irreducible structures of the Universe and recognize them as the
different facets of the one and the same quantum Universe.
Physical picture is interesting and non-reducible to the existing
interpretations. As a particularly interesting topic in this
context appears the 'free will' topic of current interest in the
interpretation of quantum theory. To this end, we arrive at the
following interesting observation. The freely chosen actions (e.g.
quantum measurements) performed by a (conscious) agent that are
still locally observable in the alternate Worlds could seem
physically unexplainable ('non-physical', 'ghostly').

\bigskip

{\it Keywords:} interpretation of quantum mechanics, quantum
structures, free will

\bigskip

{\bf 1 Introduction}

\bigskip

There is a hot ongoing debate about the interpretation of quantum
mechanical formalism; for some recent issues see e.g. (Saunders
{\it et al}., 2010; Pussey {\it et al}., 2012; Ma {\it et al}.,
2012; Vedral, 2010; Mermin, 1998; 't Hooft, 2007).

Based on some fresh looks into the quantum mechanical formalism,
here we point out a new interpretational discourse of wide-range
implications that include both consciousness as well as the issue
of free will, e.g., (Conway and Kochen, 2008; Gisin, 2010).

Our starting point is the recently re-discovered importance of the
"structure", i.e. of the decomposition into parts, subsystems, of
a composite quantum system (Dugi\' c and Jekni\' c, 2006; Dugi\' c
and Jekni\' c-Dugi\' c, 2008 ; Dugi\' c and Jekni\' c-Dugi\' c,
2012; Jekni\' c-Dugi\' c {\it et al}., 2011; Dugi\' c {\it et al},
2012; Jekni\' c-Dugi\' c {\it et al}., 2012). When applied to the
quantum Universe, this opens the new avenues not only for
interpretation but also a wide-range of implications for
describing the quantum Universe. The emerging picture is
physically interesting and mind provoking. Physical existence of
the simultaneously existing dynamical Quantum Worlds is
unquestionable. For the Universe as isolated whole, a World does
not seem more realistic than any other world. Bearing only the
common time axis and being subject to the Schr\"odinger law, these
worlds represent the parallel worlds of the completely new kind.

Our aim here is properly to describe the quantum mechanical
foundations of such, new kind of the parallel quantum worlds, and
to make a few ramifications; in this sense, we outline the {\it
bare essentials} of an emerging interpretation of quantum
mechanics. As to the later, we are particularly interested in the
possible existence of the intelligent, conscious agents in at
least some of these worlds--the world we are living in is one out
of the number of such possible worlds. If an agent in a world is
free to choose an action local to his own world, that action, if
locally observable in our world, could seem 'unexplainable' to us.
This is the reason we call this new kind of the parallel worlds
the 'Ghostly Quantum Worlds'. As a technical support of our
claims, we offer the Supplemental Information to this paper.

\bigskip

{\bf 2 Quantum structures}

\bigskip

A quantum mechanical system, $\mathcal{C}$, is defined by its
degrees of freedom, $\{x_i\}$, and by the related conjugate
momentums, $\{p_j\}$; the commutator $[x_i, p_j] =
\imath\hbar\delta_{il}$, where $\delta_{ij}$ is the so-called
Kronecker-delta. All the system's observables (measurable physical
quantities) are the analytical functions of this basic set of
observables. Nevertheless, the set of the degrees of freedom is
{\it not} unique. One can perform the different kinds and types of
the variables transformations to obtain the alternate set of the
degrees of freedom that formally define the different {\it
structures} of the composite system $\mathcal{C}$ (Dugi\' c and
Jekni\' c, 2006; Dugi\' c and Jekni\' c-Dugi\' c, 2008 ; Dugi\' c
and Jekni\' c-Dugi\' c, 2012; Jekni\' c-Dugi\' c {\it et al}.,
2011; Dugi\' c {\it et al}, 2012; Jekni\' c-Dugi\' c {\it et al}.,
2012).

To illustrate, consider a tripartite system $\mathcal{C} = 1+2+3$.
The tripartite system $\mathcal{C}$ can be presented as a
bipartite system by introducing the alternate structures, e.g.
$A+3$ or $1+B$, where the bipartite systems, $A=1+2$ and $B=2+3$.
This is formally a trivial kind of the transformations--the
particles grouping (that is essential for the quantum
teleportation protocol (Bennett {\it et al}., 1993)). The more
general and formally nontrivial transformations introduce the kind
of the degrees of freedom that are e.g. the linear combinations of
the original ones. To this end paradigmatic are the center-of-mass
($CM$) and the "internal (relative, $R$)" degrees of freedom. All
the macroscopic systems are described by these new formal
subsystems, i.e. by the bipartite structure $CM+R$. This kind of
structure is essential for the standard theory of the hydrogen
atom (Jekni\' c-Dugi\' c {\it et al}., 2012).

The variables transformations providing the different structures
of a composite system represent a general physical method.
However, only recently we have started to realize the related
subtleties appearing in the {\it quantum mechanical} context. The
following general observation is in order (Dugi\' c and Jekni\' c,
2006; Dugi\' c and Jekni\' c-Dugi\' c, 2008 ; Dugi\' c and Jekni\'
c-Dugi\' c, 2012; Jekni\' c-Dugi\' c {\it et al}., 2011; Dugi\' c
{\it et al}, 2012; Jekni\' c-Dugi\' c {\it et al}., 2012):

\smallskip

\noindent {\bf (P1)} {\it Every structure of a composite quantum
system is equally describable by the general rules and laws of
quantum mechanics}.

\smallskip

The hydrogen atom is paradigmatic (Jekni\' c-Dugi\' c {\it et
al}., 2012). The hydrogen atom can be decomposed as "electron +
proton $(e+p)$" or as (see above for notation) $CM+R$. Hydrogen
Atom (HA) is a unique composite quantum system. Its state space
and Hamiltonian as well as its quantum state are {\it unique} in
every instant in time. Nevertheless, their mathematical {\it forms
are different for the different structures}:

\begin{eqnarray}
&\nonumber& \mathcal{H}_e \otimes \mathcal{H}_p = \mathcal{H}_{HA}
= \mathcal{H}_{CM} \otimes \mathcal{H}_R, \quad [{\bf state \quad
space}]
\\&&
H_e + H_p + H_{int} = H_{HA} = H_{CM} + H_R, \quad [{\bf
Hamiltonian}] \nonumber \\&& \sum_i c_i \vert i \rangle_e \vert i
\rangle_p = \vert \Psi \rangle_{HA} = \vert \chi\rangle_{CM} \vert
n l m_l m_s\rangle_R. \quad [{\bf quantum \quad state}]
\end{eqnarray}

\noindent Thereby, inevitably, some quantum mechanical predictions
about the two structures must be different.

In Eq. (1), the left hand sides stand for the $e+p$ and the right
hand sides for the $CM+R$ structure of the hydrogen atom. Notice
that noninteraction in the $CM+R$ structure (formally provided by
the variables separation) gives rise to a tensor-product (absence
of entanglement) form of the state, where the numbers $n, l, m_l,
m_s$ denote the standard quantum numbers for the hydrogen atom
theory (Jekni\' c-Dugi\' c {\it et al}., 2012).

In this paper, of all the structures, we consider those mutually
{\it irreducible} structures. To this end, again, paradigmatic is
the  HA model, Eq. (1): (i) there is not even a common degree of
freedom for the two structures, $e+p$ and $CM+R$, and (ii) no
subsystem (e.g. $CM$ and $R$) of one structure can be decomposed
(partitioned) into the subsystems of the alternate structure (into
$e$ and $p$). Furthermore, every subsystem, $e, p, CM, R$, is
elementary--it cannot be decomposed into the more elementary
systems ("particles")--the $CM$ and $R$ systems appear as the
elementary particles for the $CM+R$ structure. The local physical
laws (interactions) are also in general different.

In many-particle systems there are many ways formally to introduce
structures that are irreducible relative to the initial one. Here,
we skip the technical details. To support intuition, we remind
that $CM+R$ is typical in the solid state physics. There, the $CM$
degrees of freedom are usually ignored--the internal vibrations
(internal energy) of a lattice are of the main importance. In
analogy with the variables separation for HA, in solid state
physics, the internal (the $R$'s) degrees of freedom are typically
transformed to provide the "normal coordinates" (i.e. the normal
vibrational modes). This chain of the transformations provides the
different, mutually irreducible structures of the one and the same
"solid body"; e.g. the phonons cannot be decomposed into the
original particles or into the "systems" defined by the $R$'s
degrees of freedom. It can be shown (cf. Supplemental
Information): the subsystems belonging to the alternate (mutually
irreducible) structures are information-theoretically separated.
Information about one subsystem (e.g. about the HA electron) is
not sufficient for describing any subsystem belonging to any
irreducible structure (e.g. the atomic $CM$ or $R$ system).
Furthermore, there is not any information flow between the
subsystems belonging to the mutually irreducible structures. The
variables transformations do not apply only to the "massive"
quantum particles. The Bogoliubov transformations (Bogoliubov,
1947) provide an example for the quantum fields. In quantum optics
one can find the transformations encompassing the variables of the
atoms and of the electromagnetic field (Stokes {\it et al}.,
2012).

Our point (P1) now reads:

\noindent {\bf (P2)} {\it Every variables transformation provides
a specific quantum mechanical description of a composite system.
If isolated ("closed"), the system is subject to the Schr\"
odinger law for every choice of the degrees of freedom
(variables). Mutually irreducible structures represent the
mutually independent and foundationally equal physical
descriptions of the composite system.}

So, physically, the structures simultaneously evolve in time, each
being described by its (local) subsystems (degrees of freedom). An
observer can in principle choose which observables of the
composite system to measure (Conway {\it et al}., 2006; Gisin,
2010). The unique quantum state of the composite system provides
unique prediction for the probability distribution for every
measurement in every instant in time. As emphasized above, the
knowledge of the probability distribution for one subsystem (e.g.
the atomic $CM$ system) is linked with the probability
distribution for a subsystem (e.g. the atomic$R$ system) belonging
to the same structure ($CM+R$). This, however, does not apply to
the probability distributions for the subsystems belonging to the
different, mutually irreducible structures. So, the different
structures represent the different facets of the (unique)
composite system.

Operationally, observation of a structure is limited by the
practical accessibility of the composite systems observables. In
practice, not all the observables are accessible in a given
physical situation. So, the choice made by the observer is
determined by the choice of the measurement apparatus and by the
general conditions the quantum system is subjected to. Fortunately
enough, these subtleties are of no importance for us before
Section 5. Below, we consider the Universe as a whole, i.e. as a
"closed" system (subject to the Schr\" odinger law) that, {\it in
principle}, cannot be observed from outside.

\bigskip

{\bf 3 New kind of the parallel quantum worlds}

\bigskip

The general assumption of our considerations is the assumption of
the universally valid and complete quantum mechanics. In our
considerations, the quantum Universe state (the universal state)
is physically real (Saunders {\it et al}., 2010; Pussey {\it et
al}., 2012). We are interested in the structures irreducible to
the structure we are a part of. To simplify notation, we denote
formally the structure we belong to by $\mathcal{W}_{\circ}$. Now
we consider the different, mutually irreducible structures of the
Universe, formally denoted as the set $\{\mathcal{W}_i\}$, whose
existence is at least formally guaranteed (Dugi\' c and Jekni\' c,
2006; Dugi\' c and Jekni\' c-Dugi\' c, 2008 ; Dugi\' c and Jekni\'
c-Dugi\' c, 2012; Jekni\' c-Dugi\' c {\it et al}., 2011; Dugi\' c
{\it et al}, 2012; Jekni\' c-Dugi\' c {\it et al}., 2012).

To keep the tracks of the nonrelativistic quantum-mechanical
description, we stick to the structures that are mutually related
by the proper canonical transformations of the degrees of freedom.
E.g. the transformations providing Eq. (1) [that are still, as
emphasized above, paradigmatic for physical considerations]:

\begin{equation}
{\vec R}_{CM} = (m_e {\vec r}_e + m_p {\vec r}_p) / m_{CM}, {\vec
\rho}_R = {\vec r}_e - {\vec r}_p,
\end{equation}

\noindent accompanied by the total mass, $m_{CM} = m_e + m_p$, and
the "reduced mass", $m_R = (m^{-1}_e + m^{-1}_p)^{-1}$, for $CM$
and $R$, respectively. The transformations Eq. (2) are invertible.
These transformations make the two structures mutually
irreducible: neither $CM$ nor $R$ can be decomposed into $e$ and
$p$, and {\it vice versa}. Measurement of e.g. ${\vec r}_p$ {\it
requires} simultaneous measurement of {\it both}, ${\vec R}_{CM}$
and ${\vec \rho}_R$. [See Supplemental Information for further
details.]

Now we emphasize:

\smallskip

\noindent {\it every structure } $\mathcal{W}_i$ {\it is a priori
no more and no less physically realistic as any other structure}
$\mathcal{W}_j$, including our own world $\mathcal{W}_{\circ}$.

\smallskip

This statement follows from the following, "obviously correct"
observations: a. the structure we are a part of is physically
realistic, and b. for the Universe as an isolated ("closed")
system, there is no {\it a priori} privileged structure. As to the
later, by  (Zanardi, 2001):

"{\it Without further physical assumption, no partition has an
ontologically superior status with respect to any other.}"

\noindent as well as by (Halliwell, Chapter 3 in (Saunders {\it et
al}., 2010)):

"{\it However, for many macroscopic systems, and in particular for
the universe as a whole, there may be no natural split into
distinguished subsystems and the rest, and another way of
identifying the naturally decoherent variables is required.}

Now, bearing in mind those structures have practically nothing in
common, we introduce a new kind of the {\it parallel quantum
worlds}:

\smallskip

\noindent {\it The Universe hosts a number of physically equal,
mutually irreducible dynamical quantum worlds. The Worlds share
the same physical time and the fundamental Schr\" odinger law,
otherwise having nothing in common.}

\smallskip

It is worth repeating: reality of these quantum worlds is a direct
corollary of the above points a and b. To this end, the central
argument is the existence of our world, which is just one out of a
set of the possible  worlds. So, the worlds we are interested in
are defined by the requirement of mutual irreducibility that
includes our world $\mathcal{W}_{\circ}$. In descriptive terms,
one can say the Worlds share the one and the same fundamental
physical {\it matter}, while their compositions (the "substances"
defined by their respective elementary particles and their
interactions) are mutually irreducible. As a consequence, one can
say there is not 'electron' or 'proton' or the hydrogen atom as
well as any other 'system' known to us in any other world. The
variables transformations for local subsystems can be performed in
every World--the variables transformations for the proton and the
electron provide the alternative structure of the hydrogen atom as
a subsystem of our World, $\mathcal{W}_{\circ}$.

Of course, one may pose the following question: whether or not
arbitrary structure (a world) can be considered physically
relevant? Without  further elaboration of this new physical
picture, we are able only to answer as follows: as long as the
structure does not raise any physical inconsistencies, there does
not seem to be any a priori reason to be rejected. Certainly,
there may be additional criteria in this respect and one such a
criterion--the {\it classicality} criterion--will be explicitly
considered below.

\bigskip

{\bf 4 Comparison with the other interpretations}

\bigskip

Below, we consider a few interpretations relevant for our
considerations.

\bigskip

{\bf 4.1. Bohmian quantum theory}

\bigskip

In Bohm's theory (Bohm, 1951), the Universe is assumed to consist
of a set of physical particles that are embedded in a quantum
field governing the particles dynamics. In every instant in time,
there is the one and unique fundamental (nonrelativistic)
structure of the Universe. In this context, the variables
transformations (Section 2) represent purely a mathematical tool,
a mathematical artifact that does not bear any physical meaning.
i.e. The alternate structures cannot be considered physically
relevant. This, of course, is in sharp contrast with our view, in
which there are not a priori reasons to reject a formally
consistent (yet irreducible relative to our world) structure of
the Universe. So, in contradistinction with (Zanardi, 2001;
Halliwell, Chapter 3 in (Saunders {\it et al}., 2010)), one can
say that Bohm's theory {\it postulates} existence of the unique
physical, ontological structure of the Universe. We believe the
Bohmian theory meets serious problems in interpreting the quantum
correlations relativity (Dugi\'  c {\it et al}., 2012). On the
other hand, the later are essential for our considerations.

\bigskip

{\bf 4.2 Everett interpretation}

\bigskip

The parallel worlds of Section 3 have nothing in common with the
Everett parallel worlds and the Multiverse interpretation of
quantum mechanics (Saunders {\it et al}., 2010). By definition,
measurement of e.g. the electron's position cannot be performed in
any other world $\mathcal{W}_i, i \neq \circ$. Electrons, protons,
hydrogen atom etc. are the subsystems exclusively in our world,
$\mathcal{W}_{\circ}$. Consequently, (by definition), there are
not the humans in any alternate World. A World $\mathcal{W}_i$ is
the subject of the Everett interpretation--one World from our
considerations defines one possible Multiverse for the Everett
interpretation. Some details in this context can be found in
(Dugi\' c and Jekni\' c-Dugi\' c, 2010). Finally, we answer the
following question: may one consider these different Multiverses
as the fundamental quantum mechanical basis for an emergent
Multiverse that is currently discussed within the new Everretian
perspective (Saunders {\it et al}., 2010)?

Whatever 'emergent' might mean, it seems necessary to assume that
there is a common 'element' for the various Multiverses. While we
do not offer a general answer, we are still able to offer an
example exhibiting the lack of such a common element.

Recently, a model of a pair of 'Brownian' particles has been
demonstrated (Dugi\' c and Jekni\' c-Dugi\' c, 2012; Jekni\'
c-Dugi\' c {\it et al}., 2011). For a composite system, $C$, one
can recognize a pair of mutually irreducible structures, $1+2$ and
$A+B$; $1+2 = C = A+B$. Formally the two models are isomorphic
thus providing the two "environments", $2$ and $B$, for the two
(one-dimensional) particles, $1$ and $A$, respectively. The two
particles, $1$ and $A$ undergo the dynamics known as the quantum
Brownian motion (QBM) (Breuer and Petruccione, 2002) (and the
references therein). So, their quantum mechanical description is
in the spirit of Section 3: every particle ($1$ and $A$) is a
subsystem in its own world ($1+2$ and $A+B$, respectively). For
the pair of particles, one can show: there does not exist any
observable, $X$, of the composite system $C$, that could
approximate measurements of any pair of observables, $A_1$ and
$B_A$, of $1$ and $A$, respectively. Thereby there is not any
alternate world describable locally by the observable $X$ that
could be 'emergent' for the two worlds, $1+2$ and $A+B$. In other
words: one cannot assume existence of any emergent property common
for both Brownian particles. Thereby we conclude: at least this
simple example (Dugi\' c and Jekni\' c-Dugi\' c, 2012) poses a
serious problem for the emergent-ism of the modern Everett theory.
The worlds defined in Section 3 are not of the Everett kind.

The possible role of {\it consciousness} within the Everett
paradigm (Lockwood, 1989; Zeh, 2000; Menskii, 2005; Mensky, 2007)
may still raise some new issues or subtleties we are not yet aware
of.

\bigskip

{\bf 4.3 Ithaca interpretation}

\bigskip

Prima facie, our quantum worlds look very much like a reminiscence
to the Mermin's Ithaca interpretation (Mermin, 1998). As long as
there is not any additional criterion for the physical relevance
of the Worlds, the two interpretations may seem indistinguishable.

However, the main criterion for the relevance of an interpretation
is simply it should reproduce what we see  in the realistic
experimental situations (Saunders {\it et al}., 2010). At this
point, as well as focusing on the irreducible structures (worlds),
we depart from the Ithaca interpretation. Actually, we introduce
the following criterion:

\smallskip

\noindent ({\bf C}) {\it Of all the Worlds introduced in Section
3, we consider physically relevant only those that bear
"classicality".}

\smallskip

"Classicality" is the very starting point in every interpretation
as one should provide the clues and possibly the rules for the
emergence of the "classical world" from the quantum substrate.
While we are still learning about the meaning of "classicality",
the above criterion (C) is clear: whatever the classicality may
mean, a World fulfilling the criterion (C) should be regarded
equally physically relevant as the World $\mathcal{W}_{\circ}$ we
live in. Existence of alternate structures supporting classicality
is virtually intractable within the modern quantum theory.
Nevertheless, there are some models supporting classicality for
some alternate structures of the model-Universe (Dugi\' c and
Jekni\' c-Dugi\' c, 2012). So, at least in principle, one may
think in the terms of the alternate, mutually irreducible quantum
worlds bearing classicality not known to any of the existing
interpretations of quantum mechanics.

\bigskip

{\bf 4.4 Summary}

\bigskip

The quantum worlds defined in Section 3 represent a new kind of
the parallel dynamical quantum worlds simultaneously hosted by the
one and the unique quantum Universe. Due to the criterion (C), of
all such quantum worlds of relevance are only those providing
classicality for at least some of their local, intrinsic
structures. The subsystems belonging to the same structure are
mutually described by the "relative states" (Everett, 1957)
description of the universal quantum state without any necessity
(Jekni\' c-Dugi\' c {\it et al}. 2011) of the "worlds branching
(splitting)" as considered within the Everett interpretation of
quantum mechanics. The subsystems belonging to the different
worlds are mutually irreducible and do not have practically
anything in common (including the elementary particles and the
local physical laws (interactions) between them). A conscious
agent cannot say which world he is a part of.

\bigskip

{\bf 5 Consciousness and free will: a speculation on the
observable effects}

\bigskip

This part is  speculative yet mind provoking. It's starting point
is quite natural: if there is not a priori reason to consider our
World privileged relative to the other Worlds, then classicality
of our world may be essentially similar (the decoherence-based) to
the classicality of any other world.

To this end, it is important to stress: Quantum mechanics is
equally valid in every World (picked up from a set of mutually
irreducible worlds). Therefore, some basic consequences of the
universally valid quantum mechanics (e.g. decoherence) may be
equally valid for at least some Worlds. Unfortunately, we do not
go beyond this general remark. E.g. we do {\it not} advocate for
any particular solution of the measurement problem. We consider
our basic findings in Section 3 as the {\it corollaries} of the
universally valid quantum mechanics, and therefore a {\it
necessary condition} for a proper solution to the measurement
problem. In our considerations, consciousness is introduced in
analogy with our-world phenomenology and intuition. At this point,
we do not dare to claim constructive role of our findings in
defining or explaining consciousness or free will as well. Rather,
{\it we proceed in analogy} with our current knowledge and
intuition--e.g., conscioussness may be emergent property of some
information-processing assemblies.

A common assumption in the philosophy of mind is that of
substrate-independence (Bostrom, 2003) (and the references
therein). In our context, it means the possibility of conscious
information processing in some alternate worlds. Unless
'intelligence', 'consciousness' and 'mind' are
substrate-dependent, there is not any reason a priori to reject
the possibility of 'intelligent' agents also in some other worlds.
Otherwise, we would be equipped with a new criterion for
distinguishing the physical relevance of the Worlds. In the
absence of such a criterion, we seem obliged to assume the
in-principle-possibility that not only our world hosts the
intelligent beings or local compositions able to emulate the
"conscious experience" as usually considered in the philosophy of
mind.

Of course, the mind-supporting composites (or 'beings') in the
alternate worlds should bear a totally different kind of
'mind'--after all (see Section 3), the physical laws underlying
the information processing are totally different (yet, owing to
the canonical transformations connecting them, fully describable
by ours) from ours. Nevertheless, as the general rules of quantum
mechanics are common for all the worlds, one can speculate about
the scientific research performed in the alternate worlds.

Interestingly, the actions performed by conscious agents in the
different Worlds provide nontrivial and {\it global} changes for
the alternate worlds. It is intriguing that, if observable locally
in an alternate world, these actions may seem
unexplainable--'ghostly'--for a local observer.

Consider a quantum measurement performed in a world by an
intelligent agent living in that world. Here, as usual, we assume
the agent is free to choose which kind of measurement to perform.
A measurement of an observable $A$ assumes the agent is capable to
act according to his free will by physically connecting the object
of measurement and the proper measurement apparatus. According to
the general rules of the quantum measurement theory, this action
induces new correlations between the object and the apparatus.
This change of the universal state is local for the agent--only
the object and the apparatus are subject to formation of the new
quantum-mechanical correlations (quantum entanglement). However,
this action is global for every alternate world: the universal
quantum state obtains a nontrivial new form bearing quantum
correlations for the subsystems belonging to that world (see
Supplemental Information).

In effect, the local action of a measurement in a world provides a
change in the universal state that is global e.g. for  our world.
While this is in principle easily mathematically presented (Dugi\'
c and Jekni\' c, 2006; Dugi\' c and Jekni\' c-Dugi\' c, 2008 ;
Dugi\' c and Jekni\' c-Dugi\' c, 2012; Jekni\' c-Dugi\' c {\it et
al}., 2011; Dugi\' c {\it et al}, 2012; Jekni\' c-Dugi\' c {\it et
al}., 2012), the physical consequences are mind provoking. If
locally observable in our world, such actions of an agent in an
alternate world would certainly look 'unphysical', 'non-causal'.
The agent is the only one aware of his actions (performed in his
own world). These actions are commonly described as a 'physical
experiment'. But in our world, there is not any reason for a
change of state of any physical object. In our world, the agent's
actions look non-spontaneous and a-causal, i.e. physically
unexpected and apparently un-explainable. For the observers in our
world not aware of the existence of the agent in an alternate
world, the free choice of the measurement made by the agent appear
simply 'ghostly'.

Of course, the possibility of the agent to make a free choice of
quantum measurement is a matter of  "free will" (Conway {\it et
al}., 2006; Gisin, 2010) that here will not be elaborated. We just
note that these 'ghostly' local effects may be absent if the
agents are short of  free will in the alternate quantum worlds. To
this end, it is important to stress: free will of the agent to
perform a measurement is essential for the effects we are
speculating about. By preparing a piece of a material and
performing a measurement, the agent performs non-spontaneous
effects that otherwise would be absent from his world. By breaking
the chain of spontaneous quantum dynamics in his world, the agent
causes the global effects for all the other worlds that cannot be
explained by the known physics in the other worlds--those effects
are not causal in the alternate worlds. Of course, there remains
the question of local observability of such global effects in an
alternate world as well as the ability of the agent(s) in the
alternate worlds to distinguish between the spontaneous and
non-spontaneous effects. Nevertheless, the physical existence of
such global effects is here for the first time pointed out.
Further elaboration and ramifications of our conclusions are under
consideration.

By emphasizing the possible role of free will, we arrive at a
position analogous to the positions based on the anthropic
principle (Barrow and Tipler, 1986; Tipler, 2003; \' Cirkovi\' c,
2002): an intelligent agent hosted in a World can nontrivially
influence dynamics of the alternate-for-him Worlds. Due to our
initial assumptions (Section 3), the Universe as a whole remains
totally indifferent regarding the local destiny of the Worlds,
which share the same the global destiny of the quantum Universe.

\bigskip

{\bf 6 Discussion}

\bigskip

The starting point of our considerations is physically
un-questionable: the structural considerations are ubiquitous in
physics. Some consequences for the composite quantum systems are
only recently recognized. To this end, the contents of the section
1 through 3 appear properly established. However, there is also
some speculative parts that should be additionally considered.

It is by now a common wisdom that consciousness, mind etc. should
not be considered exceptional in a classical world. Encouraged by
the recent notion on the parallel occurrence of decoherence
(Dugi\' c and Jekni\' c-Dugi\' c, 2012; Jekni\' c-Dugi\' c {\it et
al}., 2011) and partly by  the prevailing emergentism in modern
Everett theory (Saunders {\it et al}, 2010), we dare to assume
that all the Worlds bearing classicality may in principle host
'conscious experience'.

In our considerations, "consciousness" (as well as free will) is
assumed as a data, without any attempt of explanation. A more
elaborate consciousness-based analysis (Menskii, 2005; Mensky,
2007) may probably introduce a discourse we are not currently
aware of. Bearing this and the fact that we do not offer a
solution to the measurement problem, we can say our interpretation
is in its infancy yet.

Finally, we assume that conscious agents in at least some of the
classicality-bearing worlds can be described by free will. Then
our conclusions on the global effects for the alternate worlds,
see Section 5, are physically firmly based. We finally consider
the possible consequences of the local observability of such
global changes. In effect, free will of an agent in a world
nontrivially changes the fate of all the other alternative worlds
without any apparent cause or explanation for the agents
(observers) in those worlds.

\bigskip

{\bf 7 Conclusion}

\bigskip

What we commonly call the Universe is just one out of many
possible Worlds in the herewith presented interpretation. Every
such a world has its own physics (the set of elementary particles
and their interactions) and logic we can mathematically describe
not yet fully to understand. Every such a world is composed of its
own kind of the elementary particles and the local physical
interactions. The subsystems of a World are mutually
interdependent not yet having anything in common (but the same
time and the universal Schr\" odinger law) with the subsystems
belonging to the alternate Worlds. A conscious agent cannot say
which World he lives in. All the known basic physics and its
ramifications have counterparts in the alternate worlds but the
details are not yet investigated. So, here proposed kind of the
parallel quantum worlds is not similar to those existing in the
literature. We speculate about the possible effects locally
produced by a conscious agent in a world and emphasize the global,
physically un-explainable effects for an alternative world.

\bigskip

{\bf Acknowledgement} The work on this paper is supported by
Ministry of Science Serbia under contract no 171028.

\bigskip

{\bf References}

\bigskip

Barrow  J D, Tipler F J.  The Anthropic Cosmological Principle.
Oxford University Press, New York, 1986.

Bennett C H, Brassard G, Cr\' epeau C, Jozsa R, Peres A, Wootters
W K. Teleporting an unknown quantum state via dual classical and
Einstein-Podolsky-Rosen channels. Phys. Rev. Lett. 1993; 70(13):
1895-1899.

Bogoliubov N N. On the Theory of Superfluidity. J. Phys. (USSR).
1947; 11(1): 23-32.

Bohm D. Quantum Theory. Prentice Hall, 1951

Bostrom N, Are You Living in a Computer Simulation?. Philos.
Quart. 2003; 53(211): 243-255.

Breuer H P, Petruccione F. The Theory of Open Quantum Systems.
Clarendon Press, 2002.

Conway J, Kochen S. The Strong Free Will Theorem.
http://arxiv.org/abs/0807.3286. Accessed date: July 21, 2008.

Dugi\' c M,  Jekni\' c J. What is "System": Some
Decoherence-Theory Arguments. Int. J. Theor. Phys. 2006; 45(12):
2249-2259.

Dugi\' c M, Jekni\' c-Dugi\' c J. What Is "System": The
Information-Theoretic Arguments. Int. J. Theor. Phys. 2008; 47(3):
805-813.

Dugi\' c M, Jekni\' c-Dugi\' c J. Which Multiverse?.
http://arxiv.org/abs/1004.0148. Accessed date: March 31, 2010

Dugi\' c M, Jekni\' c-Dugi\' c J. Parallel decoherence in
composite quantum systems. Pramana 2012; DOI:
10.1007/s12043-012-0296-3 (in press).

Dugi\' c M, Arsenijevi\' c M, Jekni\' c-Dugi\' c J. Quantum
Correlations Relativity for Continuous Variable Systems. Sci.
China-Phys. Mech. Astron. 2012; (accepted)

Everett H. 'Relative state' formulation of quantum mechanics. Rev.
Mod. Phys. 1957; 29(3): 454-462.

Gisin N.The Free Will Theorem, Stochastic Quantum Dynamics and
True Becoming in Relativistic Quantum Physics.
http://arxiv.org/abs/1002.1392. Accessed date: February 6, 2010.

Jekni\' c-Dugi\' c J, Dugi\' c M, Francom A. Quantum Structures of
a Model-Universe: Questioning the Everett Interpretation of
Quantum Mechanics. http://arxiv.org/abs/1109.6424. Accessed date:
September 29, 2011.

Jekni\' c-Dugi\' c J, Dugi\' c M, Francom A, Arsenijevi\' c M.
 Quantum Structures of the Hydrogen Atom.
 http://arxiv.org/abs/1204.3172. Accessed date: April 21, 2012.

Lockwood M. Mind, Brain, and the Quantum. Oxford University Press,
1989.

 Ma X S, Zotter S, Kofler J,
Ursin R, Jennewein T,  Brukner \v C, Zeilinger A. Experimental
delayed-choice entanglement swapping. Nature Phys. 2012; 8(6):
480-485.

Menskii M B. Concept of consciousness in the context of quantum
mechanics. Physics-Uspekhi 2005; 48(4): 389-409.

Mensky M B. NeuroQuantology 2007; 5(4): 363-376.

Mermin M D. The Ithaca Interpretation of Quantum Mechanics.
Pramana 1998; 51(5): 549-565.

Pusey M F, Barrett J, Rudolph T. On the reality of the quantum
state. Nature Phys. 2012; 8(6): 476-479.

Saunders S, Barrett J, Kent A, Wallace D, (eds.). Many Worlds?
Everett, Quantum Theory, and Reality.  Oxford University Press,
2010

Stokes A, Kurcz A, Spiller T P,  Beige A.  Extending the validity
range of quantum optical master equations.  Phys. Rev. A 2012;
85(5): 053805.

't Hooft G.  A mathematical theory for deterministic quantum
mechanics. J. Phys. Conf. Ser. 2007; 67(1): 012015

Tipler J F. Intelligent life in cosmology. Int. J. Astrobiol.
2003; 2(2): 141-148.

Vedral V. Decoding Reality: The Universe as Quantum Information.
Oxford University Press, 2010.

Zanardi P. Virtual Quantum Subsystems. Phys. Rev. Lett. 2001;
87(7): 077901

Zeh H D. The Problem of Conscious Observation in Quantum
Mechanical Description. Found. Phys. Lett. 2000; 13(3): 221-233

\' Cirkovi\' c M M. Anthropic Fluctuations vs. Weak Anthropic
Principle.  Found. Sci. 2002; 7(4): 453-463.

\pagebreak

{\bf Supplemental Information}

\bigskip

We borrow notation and the references list from the main text.

\bigskip

{\bf A. The canonical transformations}

\bigskip

In nonrelativistic quantum theory, the basic observables are the
position and the momentum observables, $x$ and $p$ (e.g. for the
one-dimensional system), $[x,p] = \imath \hbar$. The "system" is
defined by its Hamiltonian, $H$, which is a function of the basic
set of the observables. The unitary evolution of the "closed"
system is generated by the Hamiltonian.

The canonical transformations preserve the formalism based on the
degrees of freedom ($x$) and the Hamiltonian of the system. Every
such variables transformation redesignes the system's Hilbert
state space. e.g. For a bipartite system $1+2$, the Hilbert state
space, $\mathcal{H}$, is defined by the "tensor-product" of the
Hilbert state spaces for the subsystems, $\mathcal{H} =
\mathcal{H}_1 \otimes \mathcal{H}_2$. The alternative structure,
$A+B$ of the same composite system ($C$) gives rise to alternate
tensor-product, $\mathcal{H} = \mathcal{H}_A \otimes
\mathcal{H}_B$.

Every such structure is fully quantum-mechanically describable--if
"closed", it is subject to the same time and the same the
fundamental quantum mechanical law--the Schr\" odinger law. In
other words: quantum mechanics, {\it per se}, does not distinguish
between the different structures of any composite system $C$.

The subsystems of the irreducible such structures are "elementary"
relative to each other--the hydrogen atom's $CM$ system can not be
'broken' to release the atomic electron and the proton, and {\it
vice versa}. The physical laws (interaction) between the
subsystems belonging to the same structure are typically (but not
necessarily (Dugi\' c and Jekni\' c-Dugi\' c, 2012; Jekni\'
c-Dugi\' c {\it et al}., 2011)) different for the different
structures.

\bigskip

{\bf B. Unique Hamiltonian and state: quantum correlations
relativity}

\bigskip

According to the postulates of quantum mechanics, for a closed
quantum system $C$, the Hilbert state space, $\mathcal{H}$, the
Hamiltonian, $H$, and the system's state, $\vert \Psi\rangle$, are
{\it unique} in every instant in time.

However, for the different structures, they all change their {\it
forms}. For an example regarding the Hilbert space see above.
Regarding the Hamiltonian:

\begin{equation}
H_1 + H_2 + H_{12} = H = H_A + H_B + H_{AB}
\end{equation}

\noindent where the double subscripts denote the interaction
terms.

Recently discovered quantum correlations relativity (Dugi\' c {\it
et al}., 2012) states: a pure quantum state $\vert \Psi\rangle$,
just like the Hilbert space and the Hamiltonian of the composite
system, obtains the different forms for the different structures.
e.g. in the hydrogen atom theory (Jekni\' c-Dugi\' c {\it et al}.,
2012) (see also Eq. (1) above):

\begin{equation}
\vert \chi\rangle_{CM} \vert nlm_l m_s\rangle_R = \sum_i c_i \vert
i\rangle_e \vert i\rangle_p;
\end{equation}

\noindent the notation is the standard notation form the quantum
theory of the hydrogen atom.

This is, a separable (no correlations) pure state for one
structure ($CM+R$) typically obtains entangled form for alternate
structure ($e+p$). The atomic center-of-mass and the internal
degrees of freedom do not mutually interact (the "variables
separation") and therefore the state on the lhs of Eq. (2) is
tensor-product. However, the atomic electron and the proton are in
mutual (unavoidable Coulomb) interaction and their state is
quantum mechanically entangled.

For the Universe as a whole: the Universe is the only physical
system {\it exactly} described by the Schr\" odinger law, and
cannot be observed from outside--there is not any observer not
belonging to the Universe. While the local variables
transformations (considered in A) are possible and often
important, these do not change the conclusions referring to the
Universe as a whole.

\bigskip

{\bf C. The global consequences of a local action}

\bigskip

"Observer" is assumed a conscious agent capable of performing
experiments that need  not spontaneously occur. Every observer is
a part of a structure (of a Universe's World) and we are ourselves
a part of such a world, $\mathcal{W}_{\circ}$, we believe to be
physically realistic. Bearing in mind the above A, there is not
any reason to elieve the Worlds irreducible to each other and to
our own World are equally physically realistic.

The subsystems belonging to the same World are described by the
states that represent the Everett "relative states" (Everett,
1957)--cf. Eq. (4). The state of the atomic electron has sense
{\it if and only if} as the atomic proton's state has physical
sense. The same applies to all the subsystems belonging to the
same structure; the local variations of structure do not change
anything in this regard.

Writing down the equalities of the form Eq. (4) is a tough
mathematical task. Nevertheless, validity of such equalities, and
their generalization:

\begin{equation}
\sum_i c_i \vert i\rangle_1\vert i \rangle_2 = \sum_p d_p \vert p
\rangle_A \vert p \rangle_B
\end{equation}

\noindent [for a bipartite structures of a composite system $C$;
$1+2 = C = A+B$] is a direct corollary of the universally valid
quantum mechanics.

In general, the expressions like Eq. (2) are time dependent--all
the time there is dynamical formation of correlations and
decorrelations in the Universe. By living in one World, we believe
the processes are spontaneous (and in a sense causal),
[statistically] predictable, all but the actions performed by an
experimenter. Every experiment can be described in a simplified
form as formation of correlations between the object of
measurement ($O$) and of the measurement apparatus ($A$):

\begin{equation}
\vert \phi\rangle_{object}\vert 0 \rangle_{apparatus} \to \sum_k
\kappa_k \vert k\rangle_{object} \vert k\rangle_{apparatus}.
\end{equation}

{\it Essential} for Eq. (6) is the fact that it {\it neglects the
rest of the Universe}. i.e. The measurement described (after von
Neumann) by Eq. (6) is {\it local} in the world the experimenter
lives in.

However, and this is the point strongly to be emphasized:

\smallskip

\noindent {\it according to the Correlations Relativity} (Dugi\' c
{\it et al}., 2012){\it , practically every action local to a
World (to one structure of the Universe) is global for every
alternate World (alternate structure of the Universe).}

\smallskip

Physically, formation of local correlations in one World is
typically presented, Eq. (6), by formation (or at least change) of
the {\it global} correlations encompassing the whole of an
alternate World. e.g. Externally-induced separation of the
hydrogen atom's electron and proton, providing a tensor-product
state in the $e+p$ structure, would  lead to formation of
entanglement in the $CM+R$ structure (Dugi\' c {\it et al}.,
2012). The fact that these are the global transformations for the
hydrogen atom does not alter our observation. In the more
technical terms: the universal-state local change induced locally
(e.g. by an experimenter) in a World produces {\it global}
consequences for the correlations in every alternate world.

As the universal Hamiltonian and the universal-state  are unique
in every instant in time, there is one-to-one prescription between
the effects in all the Worlds. So, in the Universe where
everything is spontaneous, there is not a problem. However,
existence of conscious agents in a world alternate to our own
world, if equipped by free will can produce non-spontaneous
effects physically unexplainable to us--as emphasized in the main
text.

\end{document}